\documentclass[12pt,a4paper]{article}
\usepackage{amsmath,amssymb,bm,ascmac}
\usepackage[dvipdfmx]{graphicx}
\usepackage{color}
\usepackage{authblk}
\begin{document}
\title{On three-dimensional trace anomaly from holographic local RG}
\author{Ken KIKUCHI, Hiroto HOSODA, Akihiro SUZUKI}
\affil{Department of Physics, Nagoya University}
\maketitle

\newcommand{\defi}{\stackrel{\mathrm{def}}{\iff}}
\newcommand{\tdif}[2]{\frac{d#1}{d#2}}
\newcommand{\pdif}[2]{\frac{\partial #1}{\partial #2}}
\newcommand{\ddif}[2]{\frac{\delta #1}{\delta #2}}
\newcommand{\inj}{\hookrightarrow}
\newcommand{\surj}{\twoheadrightarrow}
\newcommand{\map}[1]{\stackrel{#1}{\rightarrow}}
\newcommand{\longmap}[1]{\stackrel{#1}{\longrightarrow}}
\newcommand{\dom}[1]{\mathrm{dom}\left(#1\right)}
\newcommand{\cod}[1]{\mathrm{cod}\left(#1\right)}
\newcommand{\notsubset}{\not\subset}
\newcommand{\notsupset}{\not\supset}
\newcommand{\fsl}[1]{\not\!#1}
\newcommand{\ld}[1]{\mathfrak L_{#1}}
\newcommand{\dk}[2]{\frac{d^{#1}#2}{i(2\pi)^{#1}}}
\newcommand{\dz}[1]{\frac{d#1}{2\pi i}}
\newcommand{\dv}[2]{d^D#1\sqrt{-#2}}
\newcommand{\delB}{\bm\delta_\mathrm{B}}
\newcommand{\E}{_\mathrm{E}}

\newcommand{\mh}{\hat{\mu}}
\newcommand{\nh}{\hat{\nu}}
\newcommand{\rh}{\hat{\rho}}
\newcommand{\sh}{\hat{\sigma}}
\newcommand{\mn}{\mu\nu}
\newcommand{\rs}{\rho\sigma}
\newcommand{\p}{(\phi)}
\newcommand{\hp}{(\hat\phi)}
\newcommand{\4}{_{(4)}}
\newcommand{\3}{_{(3)}}
\newcommand{\nabb}{\bm{\nabla}}
\newcommand{\nn}{\nonumber}
\newcommand{\mL}{\mathcal L}
\newcommand{\loc}{_\text{loc}}
\newcommand{\locw}[1]{_{\text{loc};#1-3}}

\begin{abstract}
Odd-dimensional quantum field theories (QFTs) can have nonzero trace anomalies if
external fields are introduced and some ingredients needed to make Lorentz scalars
with appropriate mass dimensions (or weights) are supplied. We have studied a
three-dimensional QFT and explicitly computed the trace of the stress tensor using
the holographic local renormalization group (RG). We have checked some properties
of vector beta functions and the Wess-Zumino consistency condition, however,
found the anomalies vanish on fixed points. We clarify what is responsible for the
vanishing trace anomalies.
\end{abstract}

$Introduction$ Without a doubt, symmetry plays a central role in physics. For
example, spacetime symmetries impose various conservation laws, which govern
classical physics almost completely. The Poincar\'e symmetry is one of such a
symmetry and is one of the fundamental assumptions of quantum field theories
(QFTs). However, in quantum theories, it happens that some symmetries are violated
due to quantum corrections. They are called anomalies. Thus anomalies play
important roles in quantum theories. For instance, the chiral anomaly gave human
beings an insight how many colors Nature has. In this work, we would like to study an
anomaly called the trace anomaly. Its general classification was given in \cite{DS}.
The anomaly has also been playing a significant role in QFTs. In fact, it is known that
in two and four spacetime dimensions, coefficients of some terms in the trace
anomaly can be interpreted as the `number of degrees of freedom' in theories\cite
{cthm,athm,athm2}. See \cite{fthm} for the case of three dimensions.

Being important, the trace anomaly has been calculated in many ways. In local RG, we
lift a scale parameter into a spacetime dependent function and identify the Weyl
transformations of the metric as local scale transformations. In this line, to make
the theory consistent, coupling `constants' are forced to have spacetime
dependence and are promoted to coupling `functions'. This is why the method is
called the `local' renormalization group (LRG). The Weyl variation equation describes
the RG flow and we can obtain the trace anomalies from this equation. When we use
holography, we can get this LRG equation by the formalism made first in \cite{dVV},
generalized in \cite{FMS} to the general even spacetime dimensions, and in \cite{KS}
to gauge theories. The explicit relation between LRG and holographic
Hamilton-Jacobi method was first elucidated in \cite{RSZ}. In this formalism,
considering $(d+1)$-dimensional bulk gravity theory and by regarding one direction as
a `time', we can derive equations of motion by the Hamilton-Jacobi formalism. In the
AdS/CFT correspondence, this equation, so called flow equation, can be considered
as the LRG equation of the boundary field theory. Strictly speaking, we should say
that this method is using holography. In this way, we can compute the trace
anomalies in $d$-dimensional QFTs.

It is believed that the trace anomalies trivially vanish in odd spacetime dimensions,
and the use of the method was limited to even spacetime dimensions. However, as in
even spacetime dimensions, by introducing external fields, and furthermore by
breaking parity so as to supply an ingredient (that is, the Levi-Civita tensor) to make
Lorentz scalars from odd numbers of Lorentz indices, there is no reason for the
trace anomaly to vanish. Nakayama pointed out \cite{Nak13} the possibility and wrote
down consistency conditions the three-dimensional trace anomaly should obey. In
this work, limiting our analysis to QFTs with bulk duals, we explicitly computed the
trace anomaly and, in contrast to our optimistic expectation, ended up to find the
anomaly vanishes on fixed points. An explicit computation was also done in \cite
{BMS}\footnote{We would like to appreciate Adam Schwimmer bringing the paper to
our attention and elucidating a mechanism for producing trace anomalies.}.

The organization of the paper is as follows: Firstly, we put forward our calculations
following the well-known formalism, the Hamilton-Jacobi formalism. In this formalism
the so-called flow equation has great importance. We get the trace of the stress
tensor through this equation, and at the same time the scalar and vector beta
functions. Next, we check some properties of the beta functions and the
Wess-Zumino (WZ) consistency conditions of the anomaly coefficients following
\cite{Nak13}. Finally, we conclude our analysis and clarify the reason why the trace
anomalies vanish in our situation.\newline

$Formalism$ We start with a simple extention of the bulk action in \cite{KS} by
adding the
$\theta$-term to break the parity:
\begin{align}
	\bm S&\left[\hat\gamma_{\mh\nh}(x,\tau),\hat\phi^I(x,\tau),\hat A^a_{\mh}
		(x,\tau)\right]\nn\\
	=&\int_{M_4}d^4X\sqrt{\hat\gamma}\,\Big\{V\hp-\hat R\4+\frac12L^{IJ}\hp
		\hat\gamma^{\mh\nh}\hat\nabb_{\mh}\,\hat\phi^I\hat\nabb_{\nh}\,\hat\phi^J
			+\frac14B\hp\hat F^a_{\mh\nh}\hat F^{a\mh\nh}+\frac14\Theta\epsilon\4
				^{\mh\nh\rh\sh}\hat F^a_{\mh\nh}\hat F^a_{\rh\sh}\Big\}\nn\\
	&-2\int_{\Sigma_3}d^3x\sqrt{\hat h}\,\hat K \ ,\label{bulkS}
\end{align}
where fields with hat denote off-shell fields, which are not necessarily solutions of
equations of motion. Our notations are collected in Appendix \ref{notation}.
Since we want to identify the scalar fields with coupling functions in local
renormalization group (LRG), we restrict gauge symmetries $G$ to groups with real
representations such as $SO(N)$. The scalar fields belong to some real
representation $r$, therefore we do not distinguish upper and lower indices. The
second term on the RHS is the Gibbons-Hawking term, which is needed to treat the
action in the Hamiltonian formalism and is defined on a three-dimensional
hypersurface $\Sigma_3:=\{X\in M_4|\tau=\text{const.}\}$. Here we also defined an
induced metric $\hat h_{\mu\nu}$ and determinants $\hat\gamma:=-\det
(\hat\gamma_{\mh\nh})$ and $\hat h:=-\det(\hat h_{\mu\nu})$. Following the
traditional method, i.e., by using the ADM decomposition
\begin{equation}
	ds^2=\hat\gamma_{\mh\nh}dX^{\mh}dX^{\nh}=\hat N^2(x,\tau)d\tau^2
		+\hat h_{\mu\nu}(x,\tau)[dx^\mu+\hat\lambda^\mu(x,\tau)d\tau][dx^\nu
			+\hat\lambda^\nu(x,\tau)d\tau]\ ,
\label{ADM}
\end{equation}
and defining canonical momenta
\begin{align}
	\hat\pi^{\mu\nu}:=&\pdif{\mL_4}{(\partial_\tau\hat h_{\mu\nu})}=\hat K^{\mu\nu}
		-\hat h^{\mu\nu}\hat K\ ,\label{pimn}\\
	\hat\pi^I:=&\pdif{\mL_4}{(\partial_\tau\hat\phi^I)}=\frac1{\hat N}L^{IJ}\hp\left(\hat
		\nabb_\tau\hat\phi^J-\hat\lambda^\mu\hat\nabb_\mu\hat\phi^J\right)\ ,
			\label{piI}\\
	\hat\pi^{a\mu}:=&\pdif{\mL_4}{(\partial_\tau\hat A^a_\mu)}=\frac1{\hat N^3}
		B\hp\left[\hat N^2\hat h^{\mu\nu}\hat F^a_{\tau\nu}-\hat\lambda^\nu
			\left(\hat N^2\hat h^{\rho\mu}+\hat\lambda^\rho\hat\lambda^\mu\right)
				\hat F^a_{\nu\rho}\right]-\hat N\Theta\epsilon\4^{\mu\nu\rho\tau}
					\hat F^a_{\nu\rho}\nn\\
	&\hspace{45pt}=\frac1{\hat N}B\hp\left[\hat h^{\mu\nu}\hat F^a_{\tau\nu}
		-\hat\lambda^\nu\hat h^{\rho\mu}\hat F^a_{\nu\rho}\right]-\Theta\epsilon\3
				^{\mu\nu\rho}\hat F^a_{\nu\rho}\ ,\label{piam}
\end{align}
where\ $\bm S=\int d^3xd\tau\sqrt{\hat h}\mathcal{L}_4+(GH)$, one arrives at the first-order
action:
\begin{align}
	&\bm S\left[\hat h_{\mu\nu},\hat\phi^I,\hat A^a_\mu,\hat A^a_\tau,\hat N,
		\hat\lambda^\mu;(x,\tau)\right]\nn\\
	&=\int d^3xd\tau\sqrt{\hat h}\Bigg\{\hat\pi^{\mu\nu}\partial_\tau\hat h
		_{\mu\nu}+\hat\pi^I\partial_\tau\hat\phi^I+\hat\pi^{a\mu}\partial_\tau\hat A^a
			_\mu\nn\\
	&+\hat N\left[\frac12\hat\pi^2-\hat\pi_{\mu\nu}^2-\frac12L^{IJ}\hp\hat\pi^I
		\hat\pi^J-\frac1{2B\hp}\hat h^{\mu\nu}\hat\pi^a_{\mu}\hat\pi^a_\nu
			-\frac\Theta{B\hp}\epsilon\3^{\mu\nu\rho}\hat\pi^a_\mu\hat F^a
				_{\nu\rho}\right.\nn\\
	&\hspace{50pt}\left.+V\hp-\hat R_{(3)}+\frac12L^{IJ}\hp\hat h^{\mu\nu}
		\hat\nabb_\mu\hat\phi^I\hat\nabb_\nu\hat\phi^J+\left(\frac14B\hp
			+\frac{\Theta^2}{B\hp}\right)\hat F^a_{\mn}\hat F^{a\mn}\right]\nn\\
	&+\hat\lambda^\mu\left[2\hat\nabla^\nu\hat\pi_{\mu\nu}
		-\hat\pi^I\hat\nabb_\mu\hat\phi^I-\hat F^a_{\mu\nu}\hat\pi^{a\nu}\right]
			\nn\\
	&+\hat A^a_\tau\left[\hat\nabb_{\mu}\hat\pi^{a\mu}-(iT^a\hat\phi)^I\hat\pi^I
		\right]\nn\\
	&+(\text{GH term})\ .\label{S1st}
\end{align}
As one notices at once, the action do not contain $\tau$ derivatives of $\hat N,
\hat\lambda^\mu$ and $\hat A^a_\tau$, thus these fields are auxiliary fields, and
their equations of motion yield the first-class constraints
\begin{align}
	\hat H:=&\frac1{\sqrt{\hat h}}\ddif{\bm S}{\hat N}\nn\\
	=&\frac12\hat\pi^2-\hat\pi_{\mu\nu}^2-\frac12L^{IJ}\hp\hat\pi^I\hat\pi^J-\frac1{2B\hp}\hat h^{\mu\nu}\hat\pi^a_{\mu}\hat\pi^a_\nu-\frac\Theta{B\hp}\epsilon\3^{\mu\nu\rho}\hat\pi^a_\mu\hat F^a_{\nu\rho}\nn\\
	&\hspace{0pt}+V\hp-\hat R_{(3)}+\frac12L^{IJ}\hp\hat h^{\mu\nu}\hat\nabb_\mu\hat\phi^I\hat\nabb_\nu\hat\phi^J+\left(\frac14B\hp+\frac{\Theta^2}{B\hp}\right)\hat F^a_{\mn}\hat F^{a\mn}\approx0\ ,\label{Hcon}\\
	\hat P_\mu:=&\frac1{\sqrt{\hat h}}\ddif{\bm S}{\hat\lambda^\mu}=2\hat\nabla^\nu\hat\pi_{\mu\nu}-\hat\pi^I\hat\nabb_\mu\hat\phi^I-\hat F^a_{\mu\nu}\hat\pi^{a\nu}
\approx0\ ,\label{Mcon}\\
	\hat G^a:=&\frac1{\sqrt{\hat h}}\ddif{\bm S}{\hat A^a_\tau}=\hat\nabb_{\mu}\hat\pi^{a\mu}-(iT^a\hat\phi)^I\hat\pi^I\approx0\ .\label{Gcon}
\end{align}
(\ref{Hcon}) and (\ref{Mcon}) are Hamiltonian and momentum constraints,
respectively, which ensure `time' translation invariance and three-dimensional
diffeomorphism invariance, respectively. (\ref{Gcon}) is nothing but the Gauss's law
and it guarantees the gauge invariance of the system.

Solving the equations of motion with Dirichlet boundary conditions at $\tau=\tau_0$
\[ \bar h_{\mu\nu}(x,\tau=\tau_0)=h(x)\ ,\quad\bar\phi^I(x,\tau=\tau_0)=\phi^I(x)\ ,
	\quad\bar A^a_\mu(x,\tau=\tau_0)=A^a_\mu(x,\tau=\tau_0)\ , \]
where fields with bar indicates on-shell fields, one attains an on-shell action:
\begin{align}
	S[h_{\mu\nu}(x),\phi^I(x),A^a_\mu(x);\tau_0]:=&\bm S[\hat h=\bar h,\hat\phi=\bar\phi,\hat A=\bar A;(x,\tau_0)]\nn\\
	=&\int d^3x\int_{\tau_0}^\infty d\tau\sqrt{\bar h}\Bigg\{\bar\pi^{\mu\nu}\partial_\tau\bar h_{\mu\nu}+\bar\pi^I\partial_\tau\bar\phi^I+\bar\pi^{a\mu}\partial_\tau\bar A^a_\mu\Bigg\}\ .
\label{onshellS}
\end{align}
Its variation
\begin{equation}
	\delta S[h(x),\phi(x),A(x);\tau_0]=-\int d^3x\sqrt h\Big\{\bar\pi^{\mu\nu}
		(x,\tau_0)\delta h_{\mu\nu}(x)+\bar\pi^I(x,\tau_0)\delta\phi^I(x)+\bar\pi^{a\mu}
			(x,\tau_0)\delta A^a_\mu(x)\Big\}\label{delS}
\end{equation}
yields Hamilton-Jacobi (HJ) equations:
\begin{equation}
	\bar\pi^{\mu\nu}(x,\tau_0)=-\frac1{\sqrt h}\ddif S{h_{\mu\nu}(x)}\ ,\quad
		\bar\pi^I(x,\tau_0)=-\frac1{\sqrt h}\ddif S{\phi^I(x)}\ ,\quad\bar\pi^{a\mu}
			(x,\tau_0)=-\frac1{\sqrt h}\ddif S{A^a_\mu(x)}\ ,\quad\pdif S{\tau_0}=0\ .
\label{HJeq}
\end{equation}
Substituting the HJ equations into the Himiltonian constraint (\ref{Hcon}), we arrive
at the flow equation
\begin{equation}
	\{S,S\}(x)=\mL_3(x)\label{flow}
\end{equation}
where
\begin{align}
	\{S,S\}:=&\left(\frac1{\sqrt h}\right)^2\left[-\frac12\left(h_{\mn}\ddif S{h_{\mn}}\right)^2+\left(\ddif S{h_{\mn}}\right)^2+\frac12L^{IJ}\p\ddif S{\phi^I}\ddif S{\phi^J}\right.\nn\\
	&\hspace{80pt}\left.+\frac1{2B\p}h_{\mn}\ddif S{A^a_\mu}\ddif S{A^a_\nu}-\frac\Theta{B\p}\sqrt h\epsilon\3^{\mn\rho}\ddif S{A^{a\mu}}F^a_{\nu\rho}\right]
\label{SS}
\end{align}
and
\begin{equation}
	\mL_3:=V\p-R\3+\frac12L^{IJ}\p\nabb^\mu\phi^I\nabb_\mu\phi^J
		+\left(\frac14B\p+\frac{\Theta^2}{B\p}\right)F^a_{\mn}F^{a\mn}\ .
\label{L3}
\end{equation}
The other constraints\footnote{Some consequences of these constraints are
collected in Appendix \ref{constraints}.}, i.e., the momentum constraint and the
Gauss's law can be used to show that three-dimensional diffeomorphism and gauge
invariance are realized. In fact, the Gauss's law constraint (\ref{Gcon}) and the
Hamilton-Jacobi equations give
\begin{align}
0&=\int d^dx\sqrt{h}\,\alpha^a\left(\nabb_\mu\pi^{a\mu}-(iT^a\phi)^I\pi^I\right)
\nn\\
&=\int d^dx\left\{\nabb_\mu\alpha^a\ddif S{A^a_\mu}+\alpha^a(iT^a\phi)^I\ddif S
	{\phi^I}\right\}\label{Gaussopid}\\
&=\int d^dx\left(\delta_\alpha^{\rm gauge} A^a_\mu\ddif S{A^a_\mu}
+\delta_\alpha^{\rm gauge}\phi^I\ddif S{\phi^I}\right)
=\delta^\mathrm{gauge}_\alpha S\ .\label{Gausslaw}
 \end{align}
Here,
\begin{align}
\delta^{\rm gauge}_\alpha A^a_\mu
:=\nabb_\mu\alpha^a\equiv\nabla_\mu\alpha^a+{f^a}_{bc}
A^b_\mu\alpha^c \ ,~~
\delta^{\rm gauge}_\alpha \phi^I:=
\alpha^a(iT^a\phi)^I \ ,
\end{align}
denote an infinitesimal gauge transformation. {}Further, the momentum constraint
(\ref{Mcon}) and the Hamilton-Jacobi equations lead to
\begin{align}
0=&\int d^dx\sqrt{h}\,\epsilon^\mu\left(2\nabla^{\nu}\pi_{\mu\nu}
-\pi^I\nabb_\mu\,\phi^I-F^a_{\mu\nu}\pi^{a\nu}\right)
\nn\\
=&\int d^dx\left\{(\nabla_\mu\epsilon_\nu+\nabla_\nu\epsilon_\mu)
\ddif S{h_{\mu\nu}}+\epsilon^\mu \nabb_\mu\,\phi^I\ddif S{\phi^I}
+\epsilon^\mu F^a_{\mu\nu}\ddif S{A^a_\nu}\right\}
\nn\\
=&\delta_\epsilon S
-\int d^dx\sqrt{h}\,\epsilon^\mu A^a_\mu\left\{{\nabb}_{\nu}\pi^{a\nu}
-(iT^a\phi)^I\pi^I\right\} \ .
\label{eAgauss}
\end{align}
Here,
\begin{align}
\delta_\epsilon\phi^I:=\mL_\epsilon\phi^I
\equiv\epsilon^\mu\partial_\mu\phi^I\ ,\quad
\delta_\epsilon A^a_\mu:=\mL_\epsilon A^a_\mu\equiv\epsilon^\nu\partial_\nu A^a
_\mu+\partial_\mu\epsilon^\nu A^a_\nu \ ,\quad
\delta_\epsilon h_{\mu\nu}:=\mL_\epsilon h_{\mu\nu}
\equiv\nabla_\mu\epsilon_\nu+\nabla_\nu\epsilon_\mu \ ,
\end{align}
are Lie derivatives with respect to three-dimensional diffeomorphism.
Noting that the second term in (\ref{eAgauss}) vanishes
because of (\ref{Gcon}) implies invariance of the on-shell action
under three-dimensional diffeomorphism.

Separate the action into local and non-local parts:
\begin{equation}
	\frac1{2\kappa_4^2}S[h,\phi,A]\equiv\frac1{2\kappa_4^2}S\loc[h,\phi,A]
		-\Gamma[h,\phi,A]\ .\label{SSG}
\end{equation}
Furthermore, so as to study the flow equation systematically, we employ the
derivative expansion by assigning an additive number called weight as in a table below:
\begin{table}[h]
\begin{center}
\begin{tabular}{c|c}
elements&weight $w$\\\hline
$h_{\mu\nu}(x),\phi^I(x),\Gamma[h,\phi,A]$&0\\\hline
$\partial_\mu,A^a_\mu(x)$&1\\\hline
$R,R_{\mu\nu},R_{\mu\nu\rho\sigma},\partial^2,\ddif{}{A^a_\mu(x)},\dots$&2\\\hline
$\ddif{}{h_{\mu\nu}(x)},\ddif{}{\phi^I(x)}$&$3$\\
\end{tabular}
\end{center}
\caption{assignment of weights}
\end{table}

\[ S\loc[h,\phi,A]=\int d^3x\sqrt h\mL\loc=\int d^3x\sqrt h\sum_{w=0,2,3,\dots}
	[\mL\loc]_w \]

We parametrize the local part as below:
\begin{align}
	[\mL\loc]_0&=W\p\ ,\label{Lloc0}\\
	[\mL\loc]_2&=-\Phi\p R\3+\frac12M^{IJ}\p\nabb^\mu\phi^I
		\nabb_\mu\phi^J\ ,\label{Lloc2}\\
	[\mL\loc]_3&=\epsilon\3^{\mn\rho}D^{IJK}\p\nabb_\mu\phi^I\nabb_\nu\phi^J
		\nabb_\rho\phi^K+\epsilon\3^{\mn\rho}E^I\p(F_{\mn})^{IJ}\nabb_\rho\phi^J\nn
			\\
	&+\epsilon\3^{\mn\rho}\frac{k_\text{CS}}{4\pi}\text{tr}\left(A_\mu\partial_\nu
		A_\rho+\frac23A_\mu A_\nu A_\rho\right)\ .\label{Lloc3}
\end{align}
In order to respect flavour symmetry, $E^I$, for example, should belong to some
representation $r$ to which also $\phi^I$ belongs. Thus it must have a form
\begin{equation}
	E^I\Big(\phi(x)\Big)\equiv\phi^I(x)E\Big(\phi(x)\Big)\label{EI}
\end{equation}
with $E\p$ a flavour singlet. Similarly, if the gauge group $G$ has an antisymmetric
three-tensor, a form
\begin{equation}
	D^{IJK}\Big(\phi(x)\Big)\equiv\epsilon^{IJK}D\Big(\phi(x)\Big)\label{DIJK}
\end{equation}
with $D\p$ in flavour singlet makes the first term of (\ref{Lloc3}) $G$ invariant, and
the term is allowed just in the case.

We also define
\begin{equation}
	S_{\text{loc};w-3}:=\int d^3x\sqrt h[\mL\loc]_w\ .\label{Sw}
\end{equation}

Using the parametrization, the flow equation (\ref{flow}) is decomposed as follows:

\hspace{-17.5pt}$w=0:$
\begin{equation}
	V\p=-\frac38W^2\p +\frac12L^{IJ}\p\partial^IW\p\partial^JW\p\ ,\label{f0}
\end{equation}
$w=2:$
\begin{align}
	-1&=\frac14W\p-L^{IJ}\p\partial^IW\p\partial^J\Phi\p\ ,\label{f2-1}\\
	\frac12L^{IJ}\p&=-\frac18W\p M^{IJ}\p-L^{KL}\p\partial^KW\p\Gamma^{L;IJ}\p\nn\\
	&~~~~~~~-W\p\partial^I\partial^J\Phi\p-\frac1{2B\p}M^{IK}\p M^{JL}\p(T^a\phi)^K(T^a\phi)^L\ ,\label{f2-2}\\
	0&=W\p\partial^K\Phi\p+L^{IJ}\p\partial^IW\p M^{JK}\p\ ,\label{f2-3}
\end{align}
$w=3:$
\begin{align}
	0&=\left(\frac1{\sqrt h}\right)^2\Bigg\{2\kappa_4^2\left(h_{\rs}\ddif{S\locw0}{h_{\rs}}\right)h_{\mn}\ddif{}{h_{\mn}}\left(\Gamma-\frac1{2\kappa_4^2}S\locw3\right)-4\kappa_4^2\ddif{S\locw0}{h^{\mn}}\ddif{}{h_{\mn}}\left(\Gamma-\frac1{2\kappa_4^2}S\locw3\right)\nn\\
	&~~~~~~~~~~~~~~~~~-2\kappa_4^2L^{IJ}\p\ddif{S\locw0}{\phi^I}\ddif{}{\phi^J}\left(\Gamma-\frac1{2\kappa_4^2}S\locw3\right)-\frac{2\kappa_4^2}{B\p}h_{\mn}\ddif{S\locw2}{A^a_\mu}\ddif{}{A^a_\nu}\left(\Gamma-\frac1{2\kappa_4^2}S\locw3\right)\nn\\
	&~~~~~~~~~~~~~~~~~-\frac\Theta{B\p}\sqrt h\epsilon^{\mn\rho}\3\ddif{S\locw2}{A^{a\mu}}F^a_{\nu\rho}
\Bigg\}\ ,\label{f3}
\end{align}
$w=4:$
\begin{equation}
	\{S,S\}_{w=4}=\left(\frac14B\p+\frac{\Theta^2}{B\p}\right)F^a_{\mn}F^{a\mn}\ .\label{f4}
\end{equation}

By defining vevs in the presence of external fields $(h,\phi,A)$ as
\begin{equation}
	\langle T^{\mn}(x)\rangle:=\frac2{\sqrt h}\ddif\Gamma{h_{\mn}(x)}\ ,\quad
		\langle O^I(x)\rangle:=\frac1{\sqrt h}\ddif\Gamma{\phi^I(x)}\ ,\quad\langle J
			^{a\mu}(x)\rangle:=\frac1{\sqrt h}\ddif\Gamma{A^a_\mu(x)}\ ,
\label{vev}
\end{equation}
(\ref{f3}) can be solved for the trace of the stress tensor:
\begin{align}
	\langle{T^\mu}_\mu\rangle&=\frac2{2\kappa_4^2}h_{\mn}\frac1{\sqrt h}\ddif{S\locw3}{h_{\mn}}+\frac4WL^{IJ}\frac1{\sqrt h}\ddif{S\locw0}{\phi^I}\frac1{\sqrt h}\langle O^{'J}\rangle+\frac4{BW}h_{\mn}\frac1{\sqrt h}\ddif{S\locw2}{A^a_\mu}\langle J^{'a\nu}\rangle\nn\\
	&~~~~~+\frac1{2\kappa_4^2}\frac{4\Theta}{BW}\epsilon^{\mn\rho}\3\frac1{\sqrt h}
		\ddif{S\locw2}{A^{a\mu}}F^a_{\nu\rho}\ ,\label{trace}
\end{align}
where $\langle\mathcal O'\rangle$ is defined as a vev of an operator $\mathcal O$
with counterterms taken into account, e.g., $\displaystyle{\langle O^{'I}\rangle:=
\frac1{\sqrt h}\ddif{}{\phi^I}\left(\Gamma-\frac1{2\kappa_4^2}S\locw3\right)}$.
This expression allows us to identify the coefficients of the vevs as beta functions:
\begin{align}
	\beta^I\p:=&-\frac4{W\p}L^{IJ}\p\frac1{\sqrt h}\ddif{S\locw0}{\phi^J}\nn\\
	=&-\frac4WL^{IJ}\partial^JW\ ,\label{sbeta}\\
	\beta^a_\mu(\phi,A)\equiv\rho^a_I\p\nabb_\mu\phi^I:=&-\frac4{B\p W\p}
		h_{\mn}\frac1{\sqrt h}\ddif{S\locw2}{A^a_\nu}\nn\\
	=&\frac4{BW}M^{IJ}(iT^a\phi)^J\nabb_\mu\phi^I\ .\label{vbeta}
\end{align}
Furthermore, the first term on RHS of (\ref{trace}) is the only origin of the term so
called `Virial current'. However, since the three-dimensional theory is topological
$\delta S\locw3/\delta h_{\mn}=0$, the term trivially vanishes, i.e., there is
no Virial current in our theory.\newline

$Analysis$ According to \cite{Nakvec}, the vector $\beta$ functions must satisfy
some properties such as (i) gradient property, (ii) compensated gauge invariance, (iii)
orthogonality, (iv) Higgs-like relation, and (v) non-renormalization condition.
Although these properties are confirmed to be satisfied in even-dimensions
\cite{KS}, one can see that they are also satisfied in three spacetime dimensions: (i)
the gradient property $\beta^a\propto\delta S_\text{loc}/\delta A^a$ is manifested
in the expression (\ref{vbeta}), (ii) the compensated gauge invariance follows trivially
since the Virial current $v$ vanishes, (iii) the orthogonality can be seen via an
explicit computation thanks to the gauge invariance of $\Phi\p$ (\ref{Phigauge})
(and (\ref{f2-3})):
\begin{equation}
	\rho^{aI}\beta^I=\frac{16}{BW}(iT^a\phi)^K\partial^K\Phi=0\ ,\label{ortho}
\end{equation}
(iv) to show the Higgs-like relations, define the local RG operator
\begin{equation}
	\Delta_{\sigma}:=\int d^3x\sigma(x)\Bigg\{2h_{\mu\nu}(x)\ddif{}{h_{\mn}(x)}+\beta^I
		[\phi(x)]\ddif{}{\phi^I(x)}+\rho^{aI}[\phi(x)]\nabb_{\mu}\phi^I(x)\ddif{}{A^a_{\mu}(x)}
			\Bigg\}\ ,\label{LRGop}
\end{equation}
and by comparing coefficients of $n$-point functions, one obtains anomalous
dimensions
\begin{equation}
	\gamma^{IJ}=-\partial^J\beta^I+\rho^{aJ}(iT^a\phi)^I \ ,\quad{\gamma^a}_b=
		\rho^{cI}\delta_{bc}(iT^a\phi)^I\ ,\label{anomdim}
\end{equation}
(v) and finally, the equivalence between vanishing vector beta function and
conservation of the current operator can be proved by case analysis using the
operator identity (\ref{gaugeop}) as in \cite{KS}.

The most general form of the trace anomaly is given by\footnote{We have employed
a slightly different notation from \cite{Nak13}.}
\begin{equation}
	\Delta_\sigma\Gamma[h,\phi,A]\Big|_\text{anomaly}=\int d^3x\sqrt h\epsilon\3
		^{\mn\rho}\sigma(x)\Big\{C_{IJK}\nabb_\mu\phi^I\nabb_\nu\phi^J\nabb_\rho
			\phi^K+C^a_IF^a_{\mn}\nabb_\rho\phi^I\Big\}\ .\label{tracegen}
\end{equation}
Comparing (\ref{tracegen}) and (\ref{traceexp}), the anomaly coefficients are
identified:
\begin{align}
	C_{IJK}&=\frac1{2\kappa_4^2}\Bigg\{-2E\rho^{a[I}(iT^a)^{JK]}-2\rho^{a[I}\partial^JE(\phi iT^a)^{K]}+\epsilon^{IJK}\beta^L\partial^LD\nn\\
	&~~~~~~~~~~~~~~~~~~~~~~~~~~~~~~~~-3D\rho^{a[I}\epsilon^{JK]L}(iT^a\phi)^L-3\partial^{[I}D\epsilon^{JK]L}\beta^L\Bigg\}\ ,\label{CIJK}\\
	C^a_I&=\frac1{2\kappa_4^2}\Bigg\{-\Theta\rho^{aI}-\frac{k_\text{CS}}{4\pi}C(r)\rho^{aI}+2E(iT^a\beta)^I+\beta^K
			\partial^KE(iT^a\phi)^I\nn\\
	&~~~~~~~~~~~-E\rho^{bI}(\phi\{T^a,T^b\}\phi)
		+\partial^IE(\phi iT^a\beta)-3D\epsilon^{IJK}(iT^a\phi)^J\beta^K\Bigg\}\ .\label{CI}
\end{align}
Using these expressios, one can show that they satisfy the WZ consistency conditions, i.e.,
\begin{align*}
	3\beta^IC_{IJK}+\rho^a_JC^a_K-\rho^a_KC^a_J&=0\ ,\\
	\beta^IC^a_I&=0\ ,
\end{align*}
by exploiting the orthogonality (\ref{ortho}) and anti-symmetry of the generators
$T^a$.\newline

$Conclusion$ In this paper we have discussed the trace anomaly in three
dimensions. We expected we could get nonzero trace anomalies even on conformal
fixed points if we break the parity symmetry, but we have eventually showed it is not
the case. Now that we have finished the explicit computation, we can easily see why
the anomaly vanishes in our situation. We know that the trace of the stress tensor
has a weight $w=3$ and the non-local action $\Gamma$ has $w=0$. Then, since
coefficients of vevs of operators are identified with beta functions, the bracket (\ref
{SS}) tells us that $\beta^I\propto\delta S\locw0/\delta\phi$ and $\beta_\mu
\propto\delta S\locw2/\delta A$. With these knowledge, let us have a closer look at
the bracket. The metric and scalar field part is a good place to start. Since functional
derivatives with these fields cancel $w=-3$ from the volume element of the local
action\footnote{Note that the weight analysis tells us that non-local action no
longer contributes to $\langle{T^\mu}_\mu\rangle$.}, just pairs of local Lagrangians
whose weights sum up to three can contribute to $\langle{T^\mu}_\mu\rangle$.
Because of the absence of a local Lagrangian with $w=1$, there is no pair with
weights $3=1+2$, and all we have is a pair $3=0+3$. The fact $\delta S\locw3/\delta
h_{\mn}=0$ kills a potential contribution from the metric part of the pair, and the
scalar field part gives the scalar beta function. Thus all contributions to $\langle{T
^\mu}_\mu\rangle$ from the metric and scalar field part of the bracket is
proportional to (scalar) beta functions. Next, let us scrutinize the gauge field part.
Since functional derivatives with $A^a_\mu$ have weights $w=2$, they do not
completely cancel $w=-3$ from the volume element, and just pairs of local
Lagrangians whose weights sum up to five can contribute to $\langle{T^\mu}_\mu
\rangle$. Then just a pair $5=2+3$ can survive, however, the contribution is again
proportional to the (vector) beta function. Therefore, all contributions are
proportional to beta functions. The absence of local Lagrangians with $w=1$ is
essential. From the above argument we have learned that one needs a term with
$w=1$ which respects Lorentz and flavour symmetries in the local action in order to
achieve nonzero trace anomalies on fixed points. Following the same analysis, one
can also see that just with the simple extension of the bulk action, one cannot get
nonzero trace anomalies in the general odd dimensions, neither, because of the
absence of local Lagrangians with odd weights.


\section*{Acknowledgements}
We are grateful to Tadakatsu Sakai for valuable discussions and useful comments on
our draft.

\makeatletter
\renewcommand{\theequation}
{\Alph{section}.\arabic{equation}}
\@addtoreset{equation}{section}
\makeatother
\appendix

\section{Notation}\label{notation}
Let us denote $[\mh\nh\rh\sh]$ the sign of a permutation $(\mh\nh\rh\sh)$,
where we define
$[012\tau]\equiv+1$\footnote{Lorentz indices of the bulk $M_{4}$ are denoted by
$\hat{\mu}$,$\hat{\nu}, ...$and those of the hypersurface by $\mu$,$\nu$,....}. Then
the Levi-Civita tensor is defined by
\[ \epsilon\4^{\mh\nh\rh\sh}=-\frac1{\sqrt{|\hat\gamma|}}[\mh\nh\rh\sh]\ . \]
In this convention, we arrive at three-dimensional expression	
\begin{equation}
	\hat N\epsilon\4^{\mu\nu\rho\tau}=\epsilon\3^{\mu\nu\rho}\ .\label{LC43}
\end{equation}

Matrix representations are given by
\begin{equation}
	(A_\mu)^{IJ}:=-A^a_\mu(iT^a)^{IJ}\ ,\quad(F_{\mn})^{IJ}:=-F^a_{\mn}(iT^a)^{IJ}\ ,
\end{equation}
and covariant derivatives are defined by
\begin{align}
\hat\nabb_{\mh}\hat\phi^I&:=\hat\nabla_{\mh}\hat\phi^I-\hat A^a_{\mh}(iT^a\hat\phi)^I\ ,\\
\nabb_{\mu}\phi^I&:=\nabla_\mu\phi^I-A^a_\mu(iT^a\phi)^I\ ,\\
\nabb_{\mu}\alpha^a&:=\nabla_{\mu}\alpha^a+f^a_{bc}A^b_{\mu}\alpha^c\ ,
\end{align}
The generators $T^a$ are normalized to yield the quadratic Casimir
\begin{align}
\text{tr}(T^aT^b)\equiv\delta^{ab}C(r)
\end{align}
for some representation $r$.

Finally, we define the Levi-Civita connection in the theory space as
\begin{align}
\Gamma^{I;JK}:= \frac12(\partial^JM^{IK} + \partial^KM^{IJ}-\partial^IM^{JK} )\ .
\end{align}

\section{Some Useful Formulae}\label{formulae}
\begin{align*}
	\ddif{S_{\locw0}}{h_{\mn}}&=\frac{1}{2}\sqrt{h}h^{\mn}W\p\ ,\\
	\ddif{S_{\locw0}}{\phi^I}&=\sqrt h\partial^IW\p\ ,\\
	\ddif{S_{\locw0}}{A^a_\mu}&=0\ ,\\
	\ddif{S_{\locw2}}{h_{\mn}}&=\sqrt h\Bigg\{\Phi\p\left(R^{\mn}-\frac12h^{\mn}R_{(3)}\right)-\nabb^{\mu}\nabb^{\nu}\Phi\p+h^{\mn}\nabb^2\Phi\p\\
	&+\frac12M^{IJ}\p\left[\frac12h^{\mn}\nabb^{\rho}\phi^I\nabb_{\rho}\phi^J-\nabb^{\mu}\phi^I\nabb^\nu\phi^J\right]\Bigg\}\ ,\\
	\ddif{S_{\locw2}}{\phi^I}&=\sqrt h\Bigg\{-\partial^I\Phi\p R_{(3)}-\Gamma^{I;JK}\p\nabb^\mu\phi^J\nabb_\mu\phi^K-M^{IJ}\p\nabb^2\phi^J\Bigg\}\ ,\\
	\ddif{S_{\locw2}}{A^a_\mu}&=-\sqrt hM^{IJ}\p\nabb^\mu\phi^I(iT^a\phi)^J\ ,\\
	\ddif{S_{\locw3}}{h_{\mn}}&=0\ ,\\
	\ddif{S_{\locw3}}{\phi^I}&=\sqrt h\epsilon\3^{\mn\rho}\Bigg\{\partial^ID^{JKL}\p\nabb_\mu\phi^J\nabb_\nu\phi^K\nabb_\rho\phi^L-3\nabb_\mu D^{IJK}\p\nabb_\nu\phi^J\nabb_\rho\phi^K\\
	&~~~~~~~~~~~~~-3D^{IJK}\p\left(F_{\mn}\phi\right)^J\nabb_\rho\phi^K+\partial^IE^J\p\Big(F_{\mn}\nabb_\rho\phi\Big)^J-\nabb_\rho E^J\p(F_{\mn})^{JI}\Bigg\}\ ,\\
	\ddif{S_{\locw3}}{A^a_{\mu}}&=\sqrt h\epsilon\3^{\mu\nu\rho}\Bigg\{-3D^{IJK}\p(iT^a\phi)^I\nabb_\nu\phi^J\nabb_\rho\phi^K
	-2\nabb_\nu\left[E^I\p(iT^a\nabb_\rho\phi)^I\right]\\
	&~~~~~~~~~~~~~~~~-E^I\p\Big(F_{\nu\rho}iT^a\phi\Big)^I-\frac{k_\text{CS}}{4\pi}C(r)F^a_{\nu\rho}\Bigg\}\ .\\
\end{align*}

\section{Consequences of first-class constraints}\label{constraints}
\begin{align}
	0&=(iT^a\phi)^I\partial^IW\p\ ,\label{Wgauge}\\
	0&=(iT^a\phi)^I\partial^I\Phi\p\ ,\label{Phigauge}\\
	0&=\nabb_\mu J^{a\mu}-(iT^a\phi)^IO^I\ ,\label{gaugeop}\\
	0&=\partial^IM^{JK}\p(iT^a\phi)^I+(iT^a)^{IK}M^{JI}+(iT^a)^{IJ}M^{IK}\ ,\label{Mgauge}\\
	\nabla_\mu W\p&=\nabb_\mu\phi^I\partial^IW\p\ ,\label{Wmom}\\
	0&=\nabla^\nu T_{\mn}-\nabb_\mu\phi^IO^I-F^a_{\mn}J^{a\nu}\ ,\label{momop}\\
	0&=\nabla_\mu\Phi\p-\nabb_\mu\phi^I\partial^I\Phi\p\ .\label{Phimom}
\end{align}

\section{Explicit form of $\langle{T^\mu}_\mu\rangle$}
Substituting some formulae in Appendix \ref{formulae}, one obtains an explicit form
of the trace of the stress tensor:
\begin{align}
	\langle{T^\mu}_\mu\rangle=&-\beta^I\langle O^I\rangle-\beta^a_\mu\langle J
		^{a\mu}\rangle-\frac\Theta{2\kappa_4^2}\epsilon^{\mn\rho}\3\beta^a_\mu F^a
			_{\nu\rho}\nn\\
	&+\frac1{2\kappa_4^2}\beta^I\frac1{\sqrt h}\ddif{S\locw3}{\phi^I}+\frac1{2\kappa_4
		^2}\beta^a_\mu\frac1{\sqrt h}\ddif{S\locw3}{A^a_\mu}\nn\\
	=&-\beta^I\langle O^I\rangle-\beta^a_\mu\langle J^{a\mu}\rangle\nn\\
	&+\frac1{2\kappa_4^2}\epsilon^{\mn\rho}\3\nabb_\mu\phi^I\nabb_\nu\phi^J
		\nabb_\rho\phi^K\Bigg\{-2E\rho^{aI}(iT^a)^{JK}-2\rho^{aI}\partial^JE(\phi iT^a)
			^K+\epsilon^{IJK}\beta^L\partial^LD\nn\\
	&~~~~~~~~~~~~~~~~~~~~~~~~~~~~~~~~~~~~~~~~~~~~~~~~~~~~~~~~~~-3D\rho^{aI}\epsilon^{JKL}(iT^a\phi)^L-3\partial^ID\epsilon^{JKL}\beta^L\Bigg\}\nn\\
	&+\frac1{2\kappa_4^2}\epsilon^{\mn\rho}\3F^a_{\mn}\nabb_\rho\phi^I\Bigg\{
		-\Theta\rho^{aI}-\frac{k_\text{CS}}{4\pi}C(r)\rho^{aI}+2E(iT^a\beta)^I+\beta^K
			\partial^KE(iT^a\phi)^I\nn\\
	&~~~~~~~~~~~~~~~~~~~~~~~~~~~~~~~-E\rho^{bI}(\phi\{T^a,T^b\}\phi)
		+\partial^IE(\phi iT^a\beta)-3D\epsilon^{IJK}(iT^a\phi)^J\beta^K\Bigg\}\ .\label{traceexp}
\end{align}

\section{Adding extra terms to the bulk action}
If one adds a term
\begin{equation}
	\frac14\int_{M_4}d^4X\sqrt{\hat\gamma}\epsilon\4^{\hat\mu\hat\nu\hat\rho\hat\sigma}H^{IJKL}(\hat\phi)\hat\nabb_{\hat\mu}\hat\phi^I\hat\nabb_{\hat\nu}\hat\phi^J\hat\nabb_{\hat\rho}\hat\phi^K\hat\nabb_{\hat\sigma}\hat\phi^L\label{additional}
\end{equation}
to the bulk action, this term disturbs canonical momentum conjugate to the scalar
field as
\begin{align}
\hat\pi'^M:=&\pdif{\mL'_4}{(\partial_\tau\hat\phi^M)}\nn \\ 
=&\hat\pi^M + \epsilon\3^{\mu\nu\rho}H^{IJKM}\hat\nabb_{\mu} \hat\phi^I \hat\nabb_{\nu}\hat\phi^J \hat\nabb_{\rho}\hat\phi^K ,\label{piI'}
\end{align}
where `$\hat\pi^M$' is the same as (\ref{piI}). Following the same calculation as in
the formalism, we can write the first-order action and arrive at first-class constraints. The new term
does not change the momentum constraint and the Gauss's law constraint, however,
it changes the Hamiltonian constraint 
\begin{align}
\hat H':=&\frac1{\sqrt{\hat h}}\ddif{\bm S'}{\hat N}\nn\\
=& -\frac12L^{IJ}\hat\pi'^I \hat\pi'^J + \epsilon\3^{\mu\nu\rho}L^{IJ}\hat\pi'^{I}H^{JKLM}\hat\nabb_{\mu} \hat\phi^K \hat\nabb_{\nu}\hat\phi^L \hat\nabb_{\rho}\hat\phi^M \nn \\
&  -\frac{1}{2}\epsilon\3^{\mu\nu\rho}\epsilon\3^{\alpha\beta\gamma} L^{IJ} H^{IKLM} H^{JK'L'M'} \hat\nabb_{\mu} \hat\phi^K \hat\nabb_{\nu}\hat\phi^L \hat\nabb_{\rho}\hat\phi^M \hat\nabb_{\alpha} \hat\phi^{K'} \hat\nabb_{\beta}\hat\phi^{L'} \hat\nabb_{\gamma}\hat\phi^{M'}+...\ ,\label{Hcon'}
\end{align}
and these changes are accompanied by modifications of the flow equation. The third
term does not contribute to the trace anomaly because the term has $w=6$. The
second term can give non-zero contribution to the $\langle{T^\mu}_\mu\rangle$,
however, since the covariant derivatives already have $w=3$, the term can enter the
trace of the stress tensor only when $\hat\pi'^I$
gives a term with $w=0$ through the Hamilton-Jacobi equation. It is possible only if
the functional derivative $\delta/\delta\phi^I$ acts on $S\locw0$, and this is
nothing but $\beta^I$. Thus the additional term (\ref{additional}) does not give
non-trivial contribution to the trace anomaly, neither.

\end{document}